\documentclass[letterpaper, 10 pt, conference]{ieeeconf}  % Comment this line out if you need a4paper

\IEEEoverridecommandlockouts                              % This command is only needed if 
\usepackage{graphicx}      % include this line if your document contains figures
\usepackage{multirow}
\let\labelindent\relax
\usepackage{enumitem}
\usepackage{tabularx}
\usepackage{placeins}
\usepackage{amsfonts}
\usepackage{amsmath}
\usepackage{accents}
\usepackage{mathrsfs}
\usepackage{amssymb, bm}
\usepackage{algorithm,algpseudocode}

\algnewcommand{\Initialize}[1]{%
  \State \textbf{Initialize:}
  \Statex \hspace*{\algorithmicindent}\parbox[t]{.8\linewidth}{\raggedright #1}
}

\newtheorem{hypp}{\sc Assumption}
\makeatother
         
\DeclareMathOperator*{\argmin}{argmin} % thin space, limits underneath in displays

 \makeatletter

\title{\LARGE \bf
On the use of Anderson acceleration in hierarchical control
}

\author{Xuan-Huy Pham$^{1,2}$, Mazen Alamir$^{1}$, Fran\c cois Bonne$^{2}$ and Patrick Bonnay$^{2}$% <-this % stops a space
\thanks{$^{1}$University of Grenoble Alpes, Gipsa-lab,(xuan-huy.pham@grenoble-inp.fr)}%
\thanks{$^{2}$ University of Grenoble Alpes, IRIG-DSBT,
 F-38000, Grenoble,France}%
}

\begin{document}

\maketitle
\thispagestyle{empty}
\pagestyle{empty}

\begin{abstract}                % Abstract of not more than 250 words.
This paper investigates the use of fixed-point Anderson acceleration method (AA) to a recently proposed hierarchical control framework. Due to its model-free property, the AA-based resulting hierarchical framework becomes more generic since no mathematical model of the subsystems at the lower layer is required at the upper coordinator layer. Numerical results are proposed to evaluate the effectiveness of this approach. The paper also presents a modified version of the original hierarchical approach that involves the AA in hierarchical control.
\end{abstract}

%%%%%%%%%%%%%%%%%%%%%%%%%%%%%%%%%%%%%%%%%%%%%%%%%%%%%%%%%%%%%%%%%%%%%%%%%%%%%%%%
\section{Introduction}
Fixed point (FP) iterations are used to solve nonlinear equations of the form $G(x) = x$. FP is attractive as it enables to solve the above equation in a derivative-free setting in which only black-box calls to the $G$ operator are involved. Therefore, the application domains of the FP method cover large set of topics including the the field of advanced numerical methods (domain decomposition \cite{marini1989relaxation,garbey2005acceleration}, multigrid \cite{washio1997krylov}), multi-physics coupling in a black-box context (fluid-structure interaction), machine learning \cite{DBLP:journals/corr/abs-1805-10638} and reinforcement learning \cite{ermis2021anderson,DBLP:journals/corr/abs-1809-09501}.

Recently, a hierarchical framework has been proposed to control a network of coupled subsystems \cite{alamir2017fixed}. This framework is structured in two distinct layers. At the lower layer, subsystems dynamics are impacted by the coupling signals coming from their neighbors. At the top layer, a coordinator attempts to optimize the overall performance by finding an optimal set-point vector that minimizes a central cost and sends its components to the corresponding subsystems. 

However, ideally, the coordinator should have no knowledge of the mathematical models that govern the subsystems. To cope with this problem, the coordinator begins by some initial guess of the coupling signals and update this guess through many iterations. In order to ensure convergence of the resulting FP iterations, \cite{alamir2017fixed} use a filter whose synthesis requires some condensed knowledge regarding the coupling matrices between subsystems. This partially violates the coordinator's local information ignorance requirements.

The contribution of this paper is to replace the mentioned filter by using the Anderson Acceleration (AA) method which uses only the previous updates in order to enhance the convergence of the FP iteration. By so doing, the coordinator completely ignores the mathematical models of the subsystems, which makes the hierarchical control framework more compatible with its initial modularity goal.

This paper is organized as follows: Section \ref{prob_formu} formulates the hierarchical control problem. Section \ref{mx_sec} recalls the previously used filter in the FP iteration's definition. Section \ref{AA_sec} introduces the Anderson Acceleration (AA) method and how it can be used to replace the previously adopted FP formualtion. The section \ref{num_res} compares the two methods through several scenarios. Section \ref{discus} proposes a new approach that uses AA in the hierarchical control framework. Finally, Section \ref{conclude} concludes the paper and gives some directions for future work.

{\bf Notation}. The following notation is extensively used in the paper. For a sequence of vector $q_{i_1}, q_{i_2}, \dots$, the following concatenation operator is used:
\begin{equation}
\underset{i \in \mathcal{I}}{\oplus}q_i : = [q_{i_1}^T,q_{i_2}^T, \dots]^T,\, \, \text{with} \quad i_1 < i_2 < \dots \in \mathcal{I}
\end{equation}
Moreover, the bold-faced notation $\bm p$ denotes the profile of a vector variable $p$ over a prediction horizon of length $N$, namely:
\begin{equation}
\bm p = [p^T(k), \dots, p^T(k+N-1)]^T \in \mathbb{R}^{N\cdot n_p}
\end{equation}
\section{Problem formulation} \label{prob_formu}
\begin{figure}[h]
 \centering
\includegraphics[width = 0.8\linewidth]{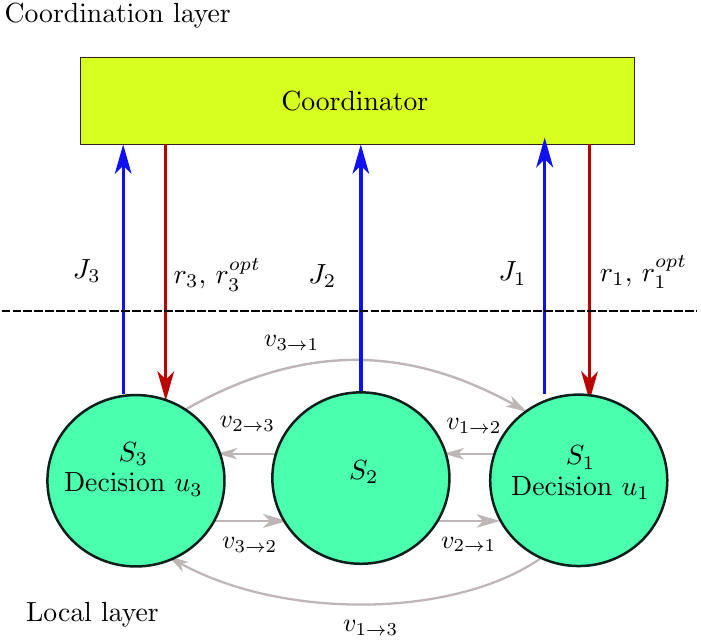}
\caption{The two layers of the hierarchical control framework.} \label{a}
 \end{figure}

Consider the control framework depicted in Fig. \ref{a}. 
The framework is separated into two layers, namely: a local layer and a coordination layer. At the local layer, there is a set of subsystems indices denoted by $\mathcal{N} = \{1,\dots, n_s\}$, which is divided into two subsets $\mathcal{N}^{ctr}$ and $\mathcal{N}^{unc}$. The subset of indices lying in $\mathcal{N}^{ctr}$ refers to subsystems that have at least one control input, while the subset $\mathcal{N}^{unc}$ contains the indices of the subsystems that do not have control (manipulated variables). 

These subsystems are coupled by the coupling variables $v_{s \rightarrow s'}$ with $s' \in \mathcal{N}_s$ where $\mathcal{N}_s$ denotes the set of subsystems' indices that affect the subsystem $S_s$. 

Let $\bm v^{in}_s$ and $\bm v^{out}_s$ indicate respectively the incoming/outgoing coupling profiles of the subsystem $S_s$. More precisely:
\begin{equation}
\bm v_s^{in} :=   \underset{s' \in \mathcal{N}_s}{\oplus} \bm v_{s' \rightarrow s} \quad;\quad 
\bm v_s^{out} := \underset{s' \vert s \in \mathcal{N}_{s'}}{\oplus} \bm v_{s \rightarrow s'}
\end{equation}
At this stage, some assumptions regarding the process occurring  at the local layer needs to be introduced:

\vskip 2mm
\hrule 
\vskip 2mm
\begin{hypp}
Each subsystem $S_s$ receives from the coordinator:
\begin{itemize}
\item a presumed incoming profile $\bm v_s^{in}$ and
\item a given individual set-point $r_s$ (required if $s\in \mathcal N^{ctr}$),
\end{itemize}   
can process an algorithm to compute what would be:
\begin{itemize}
\item Its resulting outgoing profile $\bm v_s^{out}$ and 
\item Its contribution $J_s$ to the central cost
\end{itemize}
The central cost is assumed to be of the form:
\begin{equation}
J_c(r, \bm v^{in}) := \sum_{s\in \mathcal N^{ctr}}J_s(r_s, \bm v_s^{in})+\sum_{s\in \mathcal N^{unc}}J_s(\bm v_s^{in})
\end{equation}  
where $r := \underset{s \in \mathcal{N}^{ctr}}{\oplus}r_s$ and $\bm v^{in} := \underset{s\in \mathcal N}{\oplus}\bm v_s^{in}$
\end{hypp}
\vskip 2mm
\hrule 
\vskip 2mm
Indeed, each time the coordinator sends $(r,\bm v^{in})$, this information allow the subsystems to compute their outgoing coupling profiles. In other words, there is map that depends implicitly on the current state of the subsystems (unknown to the upper coordinator level), namely:
\begin{equation}
\bm v^{out} = \bm g_{out}(r,\bm v^{in}) \label{vout_eq}
\end{equation}
It is essential to note that the elements of the outgoing coupling profile $\bm v^{out}$ are also those of the profile $\bm v^{in}$ but arranged in a different order. Certainly, both $\bm v^{in}$ and $\bm v^{out}$ are composed of all the profiles of the form $\bm v_{s \rightarrow s'}$. Thus, there is a matrix $G_{in}$ such that:
\begin{equation}
\bm v^{in} = G_{in}\cdot \bm v^{out} \label{vin_eq}
\end{equation}
Then, injecting \eqref{vout_eq} into \eqref{vin_eq}  yields:
\begin{equation}
\bm v^{in} = G_{in}\cdot \bm g_{out}(r,\bm v^{in})\label{vin_vout}
\end{equation} 
which clearly exhibits a fixed-point like equation to be solved. \\ \ \\
More precisely, it should be mentioned that the coordinator's mission is to solve an optimization problem, namely:
\begin{align}
r^{opt} = \argmin_{r} J_c(r,\bm v^{in}) \label{opti_prob} \\ 
\text{subject to:} \, \bm v^{in} = G_{in}\cdot \bm g_{out}(r,v^{in})  \label{cstr_co}
\end{align}
The constraint described by \eqref{cstr_co}, which is subsequently referred to as the coherence constraint, is satisfied in the case where the coordinator sends $\bm v^{in}$ to the subsystems, the resulting $\bm v^{out}$ is compatible with $\bm v^{in}$. Obviously, the constraint will generically be violated at the first trial that uses the arbitrary initial guess for $\bm v^{in}$. This is why a fixed-point iteration is used. The fixed-point iteration-based algorithm generally uses a stabilizing filter to enhances the convergence of the iteration \cite{alamir2017fixed} or it can be based on a residual-based iterative method to define the updated value for $\bm v^{in}$. Both methods are described in the remainder of this paper. For more details regarding the hierarchical framework studied in this paper, the reader can consult \cite{alamir2017fixed,pham2021generic,Pham2021}. This contribution focuses on the use of the AA algorithm in the specific context of the hierarchical control adopted in the references above. 

\section{Recall on Mixing method in fixed-point iterations} \label{mx_sec}
\noindent In this section, we assume that all subsystem models are linearized at an operating point $(x_s^{op}, u_s^{op})$ and that subsystems that have control inputs implement classical linear control laws such as PID-based control design, LQR or unconstrained linear MPC type. Therefore, the control profiles $\bm u_s$ is expressed as a linear (disturbance-free) equation  presented below given the current guess of the incoming coupling signal profile $\bm v_s^{in}$ at the fixed-point iteration number $\sigma$:
\begin{align} 
\tilde{\bm u}^{(\sigma)}_s := &K_s^{(x)}\cdot \tilde{x}_s(k) + K_s^{(r)}\cdot \tilde{r}_s + K_s^{(v)}\cdot \tilde{\bm v}_s^{in, (\sigma)} \label{control_prf}
\end{align}
where $\tilde{x}_s$, $\tilde{r}_s$, $\tilde{u}_s$ and $\tilde{\bm v}_s^{in, (\sigma)}$, for $s \in \mathcal{N}^{ctr}$, denote respectively the deviation of the states, set-points, control inputs and the incoming coupling profiles (at iteration $\sigma$) of the subsystem $S_s$ from their operation values. Indeed, the control profile can only be computed only with an initial guess $\tilde{\bm v}_s^{in,(\sigma)}$. 

On the other hand, the deviation of the outgoing coupling profiles can be derived from the linearized dynamics equations by using the above control profiles \eqref{control_prf}:
\begin{align}
\tilde{\bm v}^{out,(\sigma)}_{s} := \Phi^{(x)}_{s}\cdot \tilde{x}_s(k) + \Phi^{(u)}_{s}\cdot \tilde{\bm u}_s + \Phi^{(v)}_{s'}\cdot \tilde{\bm v}_s^{in,(\sigma)}  \label{vout_prd}
\end{align}
with $s\in\mathcal{N}^{ctr}$. Note that for the subsystem $S_s$ with $s \in \mathcal{N}^{unc}$, the term $\Phi^{(u)}_{s}\cdot \bm u_s$ does not exist. Note also that equation \eqref{vout_prd} is the instantiation of the general relationship \eqref{vin_vout}. Now 
Combining \eqref{vout_prd} and \eqref{control_prf}, the following equation of the coupling profile is obtained:
\begin{align}
\tilde{\bm v}_s^{out,(\sigma)} :=  \Psi^{(x)}_{s}\cdot \tilde{x}_s(k) + \Psi^{(v)}_{s} \cdot  \tilde{\bm v}_s^{in,(\sigma)} + \Psi^{(r)}_{s}\cdot \tilde{r}_s 
\end{align}
Similarly, for $s \in \mathcal{N}^{unc}$, the term $ \Psi^{(r)}_{s}\cdot \tilde{r}_s$ does not exist in the equation of $\tilde{\bm v}_s^{out,(\sigma)}$.

Then, these coupling profiles are sent to the coordinator to form the global outgoing coupling profile $\tilde{\bm v}^{out, (\sigma)}$ by concatenating all the individual $\tilde{\bm v}^{out, (\sigma)}_s$ profiles, namely:
\begin{equation}
\tilde{\bm v}^{out,(\sigma)} =  \underset{s \in \mathcal{N}}{\oplus} \tilde{\bm v}^{out,(\sigma)}_s \label{e_eq}
\end{equation}
The updated incoming coupling profiles at the next iteration is then computed by using the matrix $G_{in}$ in \eqref{vin_eq}, namely:
\begin{equation}
\hat{\bm v}^{in,(\sigma+1)} = G_{in}\cdot \tilde{\bm v}^{out,(\sigma)}  \label{compute_vin}
\end{equation}
Finally, the updated incoming coupling profiles can be written in the following condensed form:
\begin{align}
\hat{\bm v}^{in,(\sigma +1)} = \overline{M}^{(v)}\cdot \tilde{\bm v}^{in,(\sigma)}+ \overline{M}^{(x)}\cdot \tilde{x}(k) + \overline{M}^{(r)}\cdot \tilde{r} \label{vtilde_eq}
\end{align}
with
\begin{align*}
\tilde{x} = \underset{s \in \mathcal{N}}{\oplus}\tilde{x}_s,\quad
\tilde{r} = \underset{s \in \mathcal{N}^{ctr}}{\oplus}\tilde{r}_s
\end{align*}
where $\overline{M}^{(e)}$, $\overline{M}^{(x)}$, $\overline{M}^{(r)}$ and $\overline{M}^{(w)}$ are matrices coming from the matrices of the linearized models of the subsystems. 

In order to enforce the convergence of the fixed-point iteration, many  conventional mixing methods usually use some filtered version of the updated incoming coupling profile for the next iteration, namely:
\begin{equation}
\tilde{\bm v}^{in,(\sigma+1)} = (\mathbb{I}-\beta)\cdot \tilde{\bm v}^{in,(\sigma)} +\beta \cdot \hat{\bm v}^{in,(\sigma +1)}\label{filter}
\end{equation}
in which $\beta$ is often chosen to be constant.  The convergence condition for a choice of $\beta$ will be determined thereafter.

By injecting \eqref{vtilde_eq} in \eqref{filter}, we obtain:
\begin{align}
\tilde{\bm v}^{in,(\sigma +1)} =& \left[  \mathbb{I}-\beta\cdot ( \mathbb{I} -\overline{M}^{(v)}) \right] \tilde{\bm v}^{in,(\sigma)} + \beta \cdot \left[ \overline{M}^{(x)} \cdot \tilde{x}(k) + \right. \nonumber\\
& \left. \overline{M}^{(r)}\cdot \tilde{r}  + \overline{M}^{(w)}\cdot \tilde{w} \right] \label{lesMbar}
\end{align} 
This clearly shows that the convergence of the fixed-point iteration is conditioned by the spectrum radius of the matrix $\left[ \mathbb{I}-\beta\cdot ( \mathbb{I} -\overline{M}^{(v)}) \right]$. More precisely, the fixed-point iteration converges if and only if:
\begin{equation}
\rho\left(\left[ \mathbb{I}-\beta\cdot ( \mathbb{I} -\overline{M}^{(v)}) \right]\right) < 1
\end{equation}
where $\rho(Z)$ denotes the spectrum radius of the matrix $Z$, namely:
\begin{equation}
\rho(Z) := \max_i|\lambda_i(Z) |
\end{equation}
Hence, the choice of $\beta$ is crucial to the convergence of the method. In \cite{alamir2017fixed}, a more general formulation of the filtered version is proposed that takes the following form:
\begin{equation}
\tilde{\bm v}^{in,(\sigma+1)} = (\mathbb{I}-\Pi)\cdot \tilde{\bm v}^{in,(\sigma)} +\Pi \cdot \hat{\bm v}^{in,(\sigma +1)} \label{filter_new}
\end{equation}
where the scalar parameter filter $\beta$ is replaced by a matrix gain $\Pi$. 
The convergence of the fixed-point iteration is ensured if the following condition holds true:
\begin{equation}
\rho \left(\mathbb{I}-\Pi +\Pi\cdot \overline{M}^{(v)} \right) < 1
\end{equation}
This can be satisfied if the pair $\left(\mathbb{I}, [\mathbb{I}-\overline{M}^{(v)}]^T\right)$ is controllable. 
If this is the case, the appropriate matrix $\Pi$ can be obtained by using the discrete linear quadratic design tools (such as the subroutine \textsc{matlab}'s \textbf{dlqr} utility). The whole process of the mixing method is described in Algorithm \ref{alg:1}.
\begin{algorithm}
\caption{Mixing method for fixed-point iteration}\label{alg:1}
\begin{algorithmic}[1]
\Initialize{$\hat{\bm v}_s^{in,(0)}; \leftarrow 0$, $s = 1,\dots, n$; \\ $ m > 0$; $\sigma \leftarrow 0$; $\epsilon \leftarrow \infty$;}\\
Coordinator sends $r_s$ to the subsystems;
\While{$(\sigma \leq \sigma_{max})$ \textbf{and} $(\epsilon \leq \epsilon_{max})$}
\For{$s \leftarrow 1,\dots,n_s$} 
\Comment{Parallel operation performed by the subsystems}
\State Subsystem $s$ computes $\tilde{\bm v}_{s}^{out,(\sigma)}$ and sends to coordinator;
\EndFor
\State Coordinator concatenates $\tilde{\bm v}_{s}^{out,(\sigma)}$ into $\tilde{\bm v}^{out,(\sigma)}$;
\State Coordinator computes $\hat{\bm v}^{in,(\sigma+1)}$ by \eqref{compute_vin};
\State Coordinator computes the filtered version of incoming coupling profile $\tilde{\bm v}^{in,(\sigma+1)}$ by \eqref{filter};
\State Coordinator sends $\hat{\bm v}^{in,(\sigma+1)}$ to the subsystems $S_s$, for $s \in {1,\dots,n_s}$, for the next round;
\State $\sigma \leftarrow \sigma +1$;
\State $\epsilon \leftarrow \max(|\hat{\bm v}^{in,(\sigma+1)}-\hat{\bm v}^{in,(\sigma)}|,0)$;
\EndWhile
\end{algorithmic}
\end{algorithm}
Note however that the previous design and its associated convergence results hold only if the control profile takes the form of \eqref{control_prf} and the underlying linearized dynamics are representative of the true dynamics given the system's state excursion. In particular, this is generally not true when $\bm u_s$ is the solution of a constrained nonlinear MPC problem. Furthermore, the design of the filter matrix $\Pi$ needs some knowledge regarding of the underlying dynamics that is condensed in the definition of the matrices $\bar M^{(x)}$ and $\bar M^{(v)}$ that are invoked in \eqref{lesMbar}. In the next section, a method will be described that overcome the last mentioned problem while showing better convergence results in some situations.

\section{Anderson acceleration for fixed-point iteration} \label{AA_sec}
Anderson Acceleration (AA) is a residual-based iterative method that is used in order to accelerating the convergence of any fixed-point iteration. In order to introduced the principle of the AA, let us rewrite \eqref{vin_vout} as a general fixed-point equation:
\begin{equation}
\bm v^{in} = G(\bm v^{in}) \label{G_funct}
\end{equation}
AA aims to accelerate the convergence of any fixed-point iteration by only using information from the most recent $m_{\sigma}$ values $\bm v^{in,(\sigma)}$. More precisely, the AA update at the $\sigma$-{th} iteration is given by:
\begin{align}
\bm v^{in,(\sigma+1)} &= G(\bm v^{in,(\sigma)})- \sum_{j=1}^{m_\sigma}  \left[ G(\bm v^{in,(\sigma-m_\sigma+j)}) \right. \nonumber \\
& \left. - G(\bm v^{in,(\sigma-m_\sigma+j-1)}) \right]\cdot\gamma_j^{(\sigma)}
\end{align} 
Since the function $G$ is not explicitly known by the coordinator, the coordinator thus receives  the estimates  computed by the subsystems. First, the subsystems compute the outgoing coupling profile given the incoming coupling profile $\bm v^{in,(\sigma)}$, namely:
\begin{equation}
\hat{\bm v}^{out,(\sigma)}= \bm g_{out}(\bm v^{in,(\sigma)}) \label{est_vout}
\end{equation}
Then, the estimation of the incoming coupling profile can be computed by the coordinator by rearranging the elements of $\hat{\bm v}^{out,(\sigma)}$ by using matrix $G_{in}$, namely:
\begin{equation}
\hat{\bm v}^{in,(\sigma)} = G_{in}\cdot \hat{\bm v}^{out,(\sigma)} \label{est_vin}
\end{equation}
Note that by combining the equations  \eqref{est_vout} and \eqref{est_vin} we obtain the same fixed-point equation \eqref{G_funct} by defining:
\begin{equation}
G(\cdot) = G_{in}\cdot \bm g_{out}(\cdot)
\end{equation}
Let us define the residual function by:
\begin{align}
g_{\sigma}:= g(\bm v^{in,(\sigma)})&= G(\bm v^{in,(\sigma)})-\bm v^{in,(\sigma)} \nonumber\\   
& = \hat{\bm v}^{in,(\sigma)} 	- \bm v^{in,(\sigma)}
\end{align}
The updating rule becomes:
\begin{align}
&\bm v^{in,(\sigma+1)} =  \bm v^{in,(\sigma)} + g_{\sigma} - \sum_{j = 1}^{m_\sigma} \left[ (\bm v^{in,(\sigma-m_\sigma+j)} \right. \nonumber\\
&\left. - \bm v^{in,(\sigma-m_\sigma+j-1)})- ( g_{\sigma-m_\sigma +j} -g_{\sigma-m_\sigma +j-1} ) \right]\cdot\gamma_j^{(\sigma)} \label{AA_1}
\end{align} 
The parameters $\gamma_j^{(\sigma)}$ are chosen in order to minimize the distance between $g(\bm v^{in,(\sigma)})$ and the linear combination of the differences $\sum_{j=1}^{m_{\sigma}} [g_{\sigma - m_\sigma +j } - g_{\sigma - m_\sigma +j-1}]\cdot \gamma^{(\sigma)}_j$, namely:
\begin{align}
\gamma^{(\sigma)} = &\argmin_{\gamma \in \mathbb{R}^{m_\sigma}} \| g_{\sigma}- \sum_{j=1}^{m_{\sigma}}  [g_{\sigma - m_\sigma +j} \nonumber\\
&- g_{\sigma - m_\sigma +j-1}]\cdot \gamma^{(\sigma)}_j\| \label{eq_gamma}
\end{align}
By defining the matrices below:
\begin{align*}
\mathcal{V}_{\sigma} &= \left[\bm v^{in,(\sigma-m_\sigma+1)}- \bm v^{in,(\sigma-m_\sigma)}{\dots } \bm v^{in,(\sigma)}- \bm v^{in,(\sigma-1)}\right] \\
\mathcal{G}_{\sigma} &= \left[g_{\sigma-m_\sigma+1}- g_{\sigma-m_\sigma}\dots g_{\sigma}- g_{\sigma-1}\right]
\end{align*}
The equation \eqref{AA_1} becomes:
\begin{equation}
\bm v^{in,(\sigma+1)} = \bm v^{in,(\sigma)} + g_{\sigma} - ( \mathcal{V}_{\sigma}+ \mathcal{G}_\sigma)\cdot \gamma^{(\sigma)}
\end{equation}
The vector $\gamma^{(\sigma)}$ at iteration $\sigma^{th}$ in \eqref{eq_gamma} is computed by solving the following optimization problem:
\begin{equation}
\gamma^{(\sigma)} = \argmin_{\gamma \in \mathbb{R}^{m_\sigma}} \|g_\sigma - \mathcal{G}_\sigma \cdot \gamma \|^2 \label{gamma_prob}
\end{equation}
Note that periodic restarts can be included in the Anderson acceleration algorithm, meaning that the acceleration scheme is restarted periodically using only the information from the most recent iterations. Such restarting mechanism is well known in the numerical analysis literature concerning conjugate gradient and  quasi-Newton iterations to cite but few examples \cite{meyer1976convergence,Powell2009TheBA}.

In the following investigations, following the proposition made by \cite{pratapa2015restarted}, the original AA algorithm is modified to include systematic restarts instead of adaptive restarts. Specifically, at some iterations at the beginning of the algorithm, columns are added to the $\mathcal{V}_\sigma$ and $\mathcal{G}_\sigma$ matrices, while their allowed  number of columns $m_\sigma$ is incremented over  iterations. Until $m_{\sigma}$ reaches the maximum number of columns defined by $m$, the algorithm is restarted using only the one-column version of $\mathcal{V}_\sigma$ and $\mathcal{G}_\sigma$ in the next iteration and the matrices $\mathcal{V}_\sigma$ and $\mathcal{G}_\sigma$  continue to be filled in until they reach the maximum number of columns $m$. The process of building the one-column to $m$-column $\mathcal{V}_\sigma$, $\mathcal{G}_\sigma$ matrices can be considered a single "cycle", and after reaching the end of the cycle, this process is restarted. This modified AA scheme to include systematic restarts is detailed in Algorithm \ref{alg:cap}.\\

\begin{algorithm}
\caption{Anderson Acceleration with restarts. In the description of the algorithm, $g(\bm v^{in,(\sigma)}) = G(\bm v^{in,(\sigma)}-\bm v^{in, (\sigma)}$, $\Delta \bm v^{in,(i)} = \bm v^{in,(i+1)}-  \bm v^{in,(i)}$, $g_i = g(\bm v^{in,(i)})$, $\Delta g_i = g_{i+1}-g_i$, $\mathcal{V}_{i} =  \left[\Delta \bm v^{in,(i-m_\sigma)},..., \Delta \bm v^{in,(i-1)} \right]$, and $\mathcal{G}_i = \left[ \Delta g_{i-m_{\sigma}},..., \Delta g_{\sigma -1}\right]$}\label{alg:cap}
\begin{algorithmic}[1]
%\Given $n \geq 0$
%\Ensure $y = x^n$
%\State $y \gets 1$
%\State $X \gets x$
%\State $N \gets n$
%\While{$N \neq 0$}
%\If{$N$ is even}
%    \State $X \gets X \times X$
%    \State $N \gets \frac{N}{2}$  \Comment{This is a comment}
%\ElsIf{$N$ is odd}
%    \State $y \gets y \times X$
%    \State $N \gets N - 1$
%\EndIf
%\EndWhile
\Initialize{$\bm v_s^{in,(0)}; \leftarrow 0$, $s = 1,\dots, n$; \\ $ m > 0$; $\sigma \leftarrow 0$;$c \leftarrow 0$; $\epsilon \leftarrow \infty$;}\\
Coordinator sends $r_s$ to the subsystems;
\While{$(\sigma \leq \sigma_{max})$ \textbf{and} $(\epsilon \leq \epsilon_{max})$}
\For{$s \leftarrow 1,\dots,n_s$} 
\Comment{Parallel operation performed by the subsystems}
\State Subsystem $s$ computes $\hat{\bm v}_{s}^{out}$ and sends to coordinator;
\EndFor

\Comment{The operations below are performed by the coordinator}
\State Coordinator forms up $\hat{\bm v}^{out,(\sigma)} := \underset{s \in \mathcal{N}}{\oplus}\hat{\bm v}^{out,(\sigma)}_s$;
\State $m_{\sigma} = \min(m,c)$;
\State $\hat{\bm v}^{in,(\sigma)} = G_{in}\cdot \hat{\bm v}^{out,(\sigma)}$;
\State  $g_\sigma =  \hat{\bm v}^{in,(\sigma)} - \bm v^{in,(\sigma)} $;

\If{$\sigma == 0$}
\State	$\bm v^{in,(\sigma +1)} = \hat{\bm v}^{in,(\sigma)}$;
\State  $\Delta \bm v^{in,(\sigma)} = \bm v^{in,(\sigma+1)}-  \bm v^{in,(\sigma)}$;
\Else{}
\State $\Delta g_{\sigma} = g_{\sigma} - g_{\sigma-1}$;
\State $ \mathcal{G}_\sigma  =[\Delta g_{\sigma - m_\sigma},\dots, \Delta g_{\sigma - 1}]$;
\State $\mathcal{V}_\sigma = \left[\Delta \bm v^{in,(\sigma-m_\sigma)},..., \Delta \bm v^{in,(\sigma-1)} \right]$;
\State Coordinator gets $\gamma^{(\sigma)}$ by solving \eqref{gamma_prob};
\State $\bm v^{in,(\sigma+1)} = \bm v^{in,(\sigma)} + g_{\sigma} - (\mathcal{V}_{\sigma}+ \mathcal{G}_\sigma)\cdot \gamma^{(\sigma)}$;
\State $\Delta \bm v^{in,(\sigma)} = \bm v^{in,(\sigma+1)}-  \bm v^{in,(\sigma)}$;
\EndIf
\\
\If{$c == m$} \Comment{check for restart}
\State $c \leftarrow 1$;
\Else{}
\State $c \leftarrow c+1;$
\EndIf
\State $\sigma \leftarrow \sigma +1$;
\State $\epsilon \leftarrow \max(|\bm v^{in,(\sigma +1)}- \bm v^{in,(\sigma)}|,0)$;
\EndWhile
\end{algorithmic}
\end{algorithm}

\section{Numerical investigations}\label{num_res}
The hierarchical control framework has been validated in many previous works 
\cite{alamir2017fixed,pham2021generic} and \cite{Pham2021} where the relevance and the effectiveness have been assessed. The objective of this section is to validate the application of the AA method by analyzing the convergence of the iteration. 

\begin{figure}[h]
 \centering
\includegraphics[width = 0.5\linewidth]{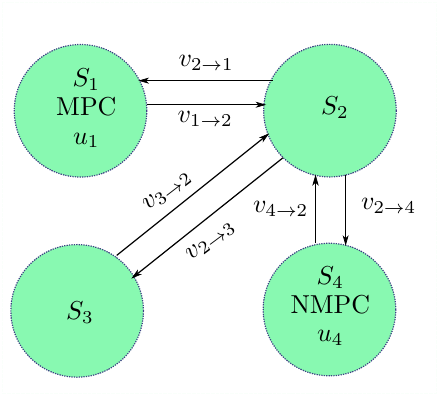}
\caption{Interconection of a network of four subsystems.} \label{4_topology}
 \end{figure}

The decomposition of the system into a a network of connected subsystems is shown in Fig.\ref{4_topology} (see the above references for more details regarding the physical signification of these subsystems). Basically, the network has four subsystems in which subsystems $S_1$ and $S_4$ have control inputs/ outputs denoted by $u_1 \in \mathbb{R}^2$, $u_4 \in \mathbb{R}$/ $y_1 \in \mathbb{R}^2$, $y_4 \in \mathbb{R}^2$. Subsystem $S_1$ also has a disturbance input denoted by $w_1 \in \mathbb{R}$. Subsystems $S_3$ and $S_4$ have no control input but are affected by their neighbors through the coupling signal $v_{s}^{in}$. Specifically, subsystems $S_1$ and $S_4$ are controlled by MPC and NMPC, respectively.

\subsection{Convergence analysis under the updating rule \eqref{lesMbar}}
\begin{figure}[h]
 \centering
\includegraphics[trim = 3.5cm 8.5cm 4cm 8.5cm,clip,width =0.5 \linewidth]{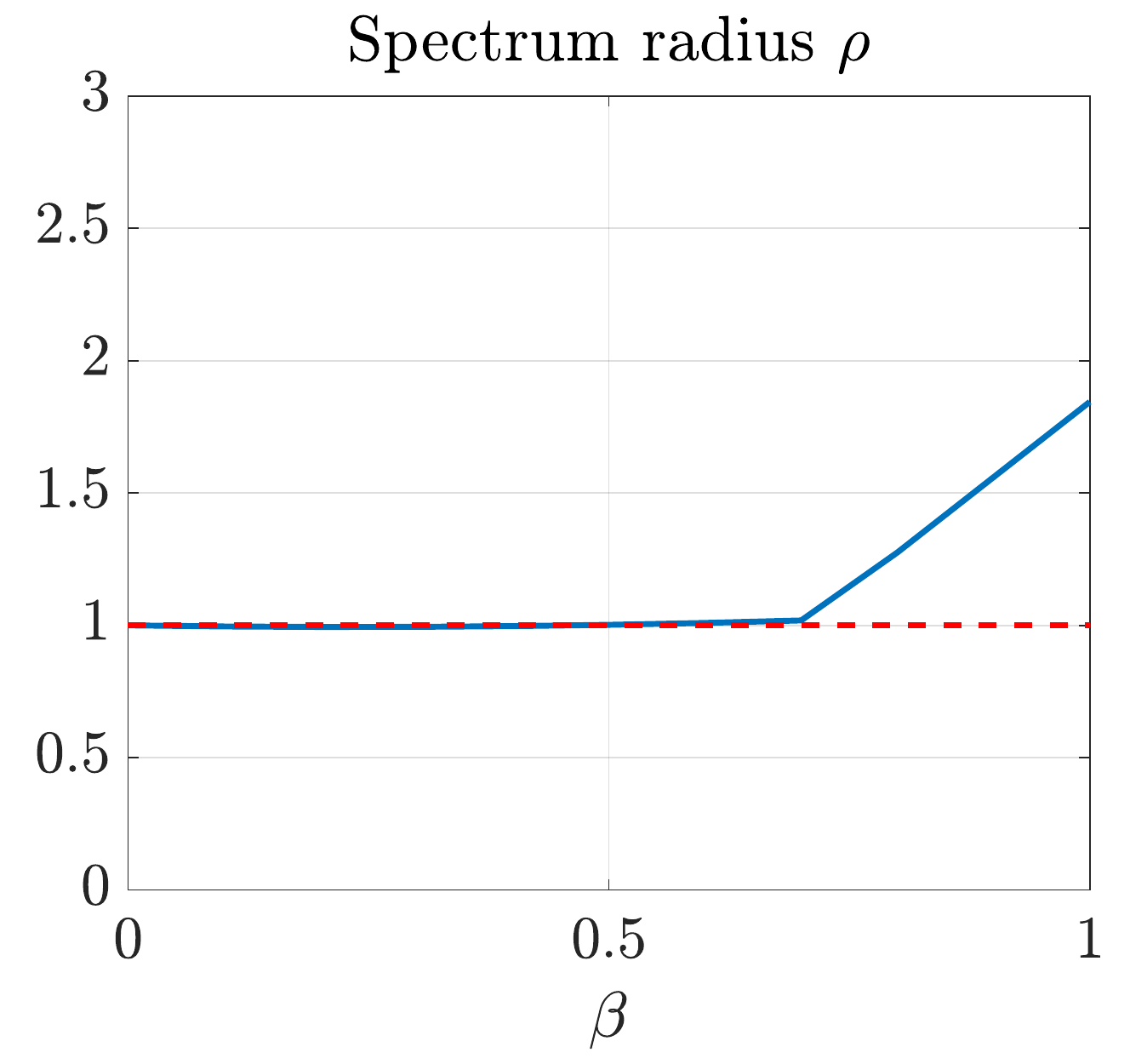}
\caption{The evolution of the spectrum when using the scalar $\beta$ to design the FP iteration update applied to the linearized equations of the subsystems.} \label{evol_spec}
 \end{figure}

Fig. \ref{evol_spec} illustrates the spectral radius of the matrix $\left[ \mathbb{I}\right.$ $\left. -\beta\cdot ( \mathbb{I} -\overline{M}^{(v)}) \right]$ for different values of $\beta$. The stability associated to the filtering law using $\beta$ is  significantly poor when almost all the choices ranging from $0$ to $1$  that makes the spectrum radius $\rho$ greater than $1$.

\subsection{Influence of the choice of memory length $m$ in AA method}
\begin{figure}[h]
 \centering
\includegraphics[trim = 3.5cm 8.5cm 4cm 9cm,clip,width =0.6 \linewidth]{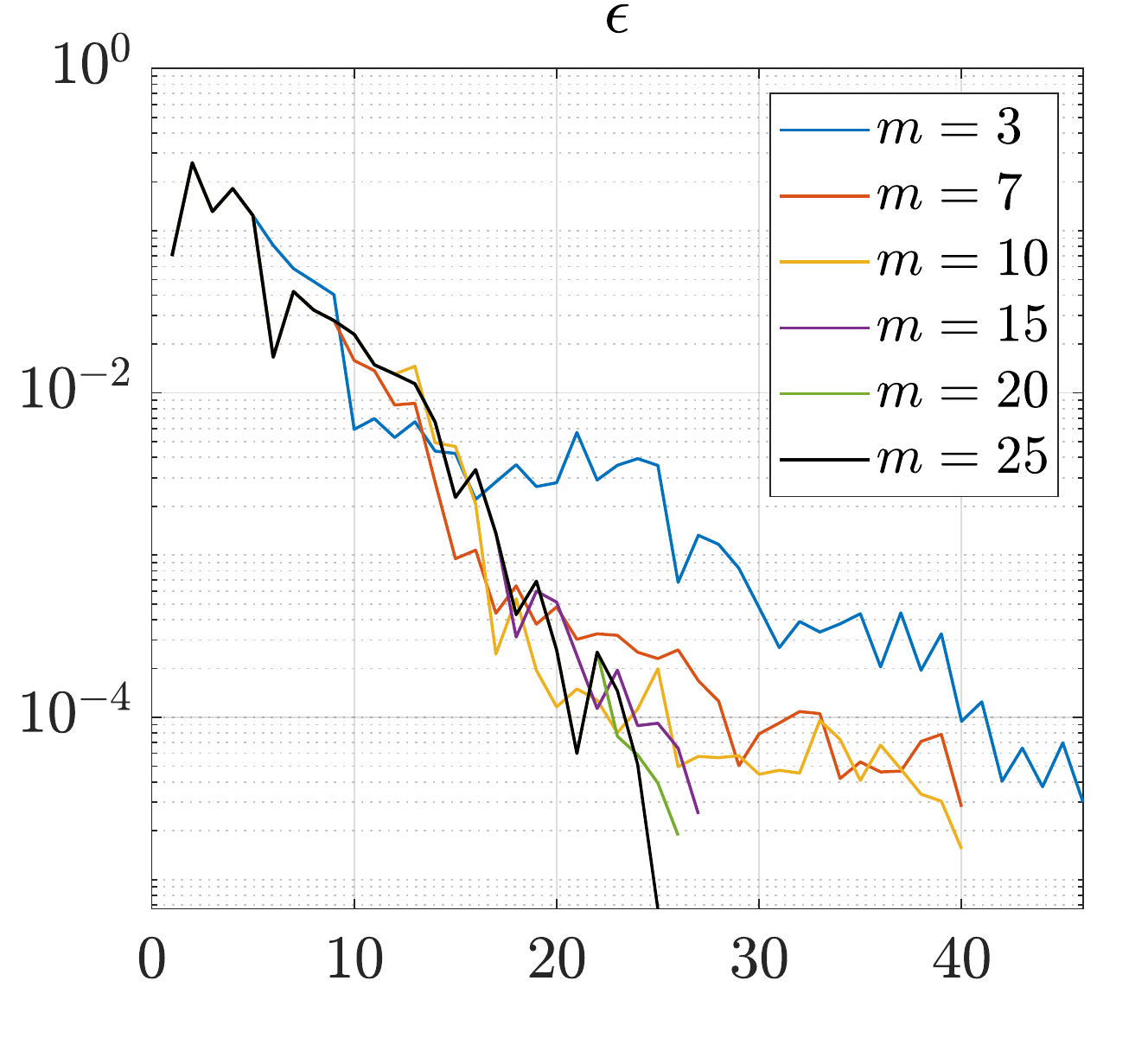}
\caption{Convergence rate of several choices of memory length $m$ in the AA method.} \label{plot_m}
 \end{figure}

Fig. \ref{plot_m} shows the evolution of convergence error for several choices of memory length $m$. This figure suggests that the use of higher $m$ generally enhances the convergence speed. 

\subsection{Comparison between advanced filter \eqref{filter_new} and AA method}
\begin{figure}[h]
 \centering
\includegraphics[trim = 3cm 8.5cm 4cm 7.5cm,clip,width =0.55 \linewidth]{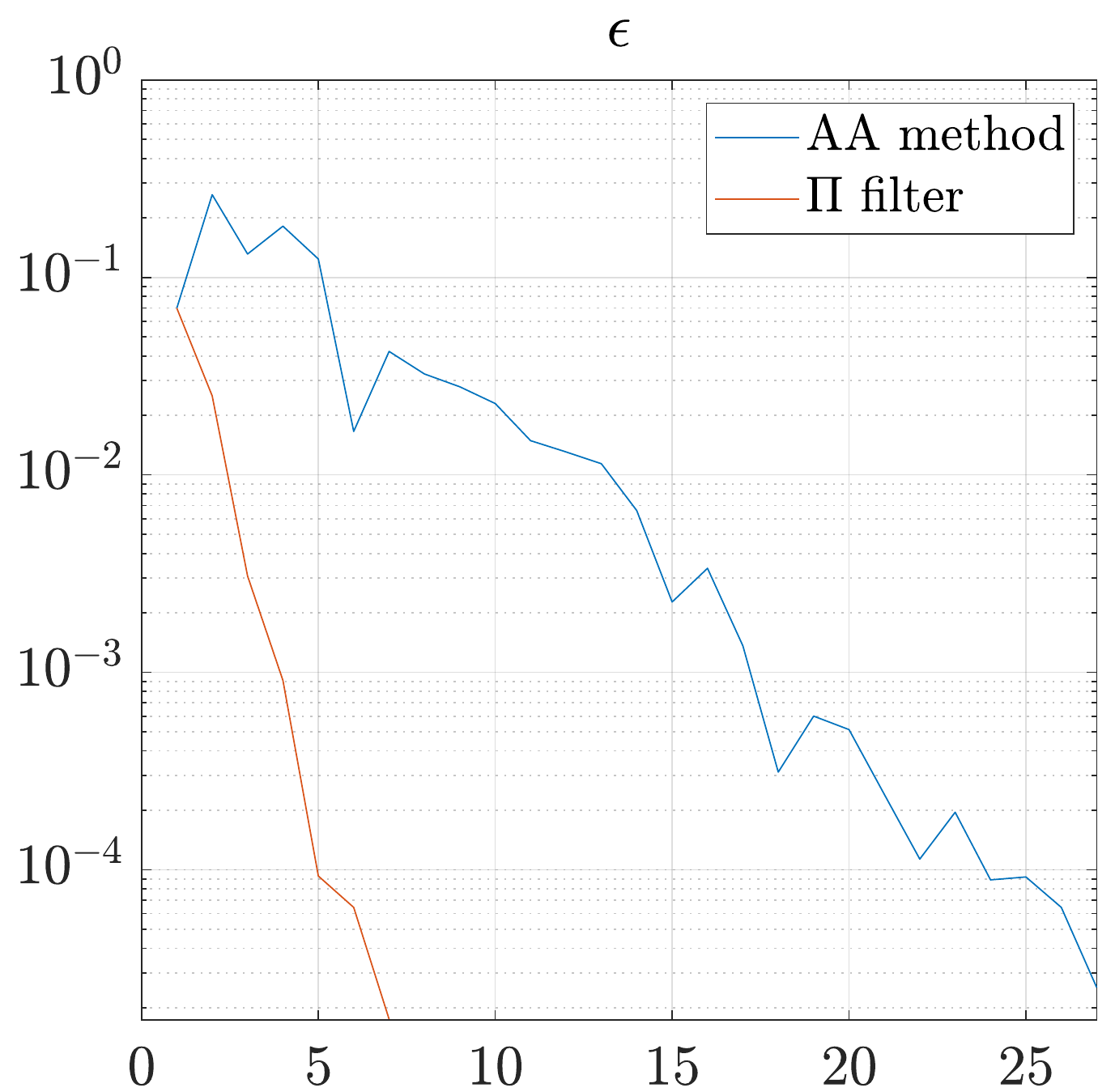}
\caption{Comparison between using the AA method and the advanced filter \eqref{filter_new} that uses the a matrix $\Pi$ that exactly corresponds to the control penalties used in the subsystems' controllers. Note that $m = 15$ is chosen for the AA method.} \label{compare_iter_1}
 \end{figure}

\begin{figure}[h]
 \centering
\includegraphics[trim = 3.5cm 8.5cm 4cm 8.5cm,clip,width =0.6 \linewidth]{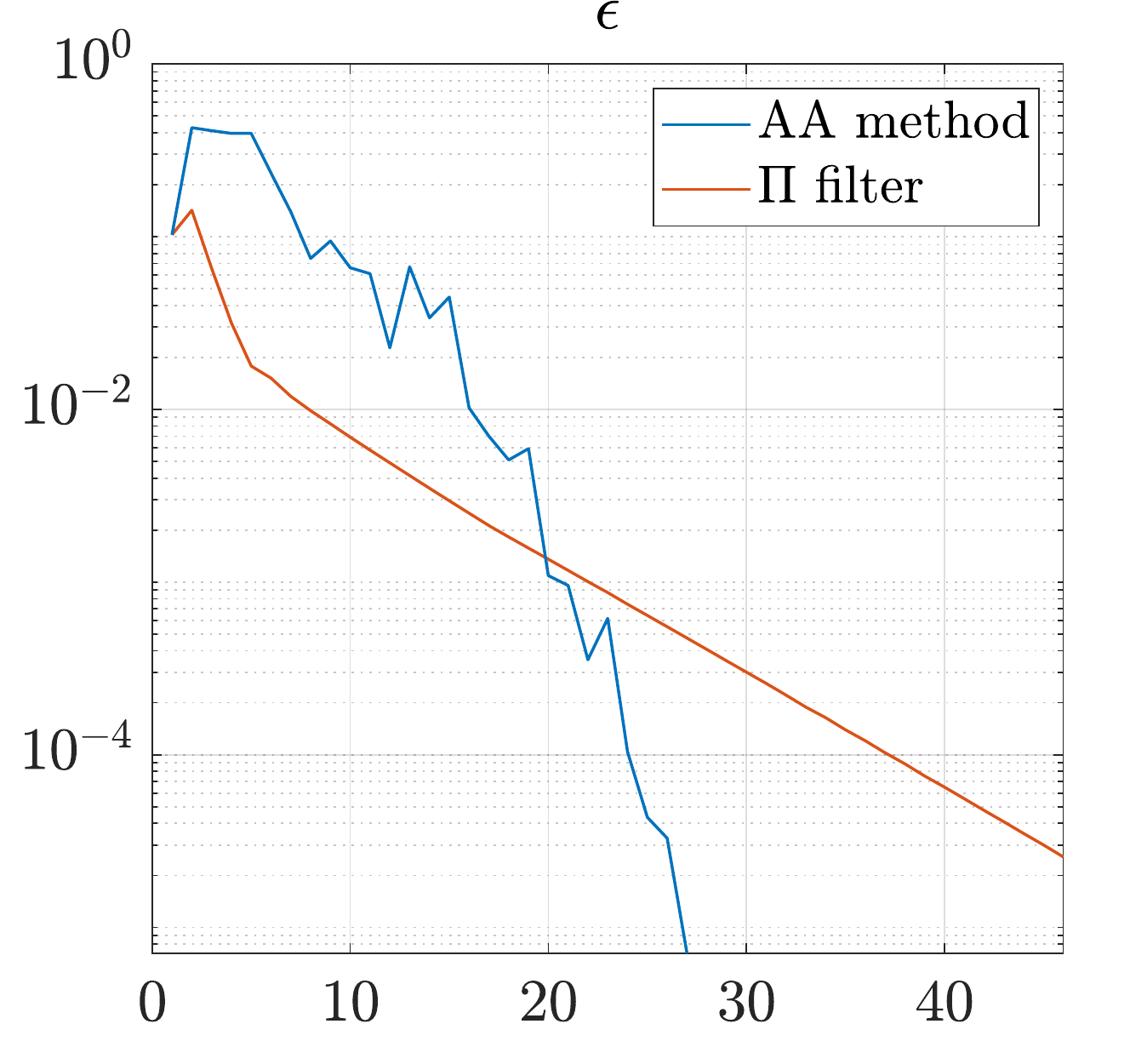}
\caption{Comparison between using the AA method and a detuned advanced filter. Note that $m = 15$ is chosen for the AA method.} \label{compare_iter_2}
 \end{figure}

Fig. \ref{compare_iter_1} compares the convergence error $\epsilon$ behavior under the  advanced filter ($\Pi$ filter) and when using the AA method for the same initial guess. It is shown the use of the advanced filter induces faster convergence. However, the key point is to note that the design of the advanced filter is specified by the control law given by \eqref{control_prf}. In other words, any changes at the control law will affect the performance of the filter, for instance, changes in the weighting matrices $Q$ and $R$ in MPC. Fig. \ref{compare_iter_2} shows a comparison of the two methods in the case where the penalty matrices used in the local MPC of $S_1$ are changed while the matrix $\Pi$ kept unchanged. This last figure shows that beyond some required accuracy, the AA might outperform the advanced filter solution. Note however that even if the advanced filter is better, its design needs a certain amount of specific knowledge while the AA framework is totally agnostic to any such a priori knowledge and can be used on the top of any pre-existing control design at the local level. 

\section{Further discussion on the application of the AA method in hierarchical control} \label{discus}
In the previous works \cite{alamir2017fixed,Pham2021,pham2021generic}, each subsystem which have decision variable $u_s$ is controlled by a local controller. Then, the coordinator tries to solve the optimization problem \eqref{opti_prob} in which the decision variable is the vector of set-points $r$, namely:
\begin{align}
r^{opt} = \argmin_{r} J_c(r,\bm v^{in})  \\ 
\text{subject to:} \, \bm v^{in} = G_{in}\cdot \bm g_{out}(r,v^{in})  
\end{align}
Keeping the same spirit of the previous works, a different hierarchical control framework can be used that keeps the assumption according to which the coordinator ignores all the mathematical models of the subsystems. This can be briefly described as follows:
\vskip 2mm
\hrule 
\vskip 2mm
\begin{hypp}
Each subsystem $S_s$ receives from the coordinator:
\begin{itemize}
\item a presumed incoming profile $\bm v_s^{in}$ and
\item a given \textbf{control profile} $\bm u_s$ (required if $s\in \mathcal N^{ctr}$),
\end{itemize}   
so that $S_s$ can process an algorithm to compute what would be:
\begin{itemize}
\item Its resulting outgoing profile $\bm v_s^{out}$ and 
\item Its contribution $J_s$ to the central cost
\end{itemize}
The central cost is assumed to be of the form:
\begin{equation}
J_c(\bm u, \bm v^{in}) := \sum_{s\in \mathcal N^{ctr}}J_s(\bm u_s, \bm v_s^{in})+\sum_{s\in \mathcal N^{unc}}J_s(\bm v_s^{in})
\end{equation}  
where $\bm u := \underset{s \in \mathcal{N}^{ctr}}{\oplus}\bm u_s$ and $\bm v^{in} := \underset{s\in \mathcal N}{\oplus}\bm v_s^{in}$
\end{hypp}
\vskip 2mm
\hrule 
\vskip 2mm

Consequently, the optimization problem that the coordinator needs to solve is redefined below:
\begin{align}
\bm u^{opt} = \argmin_{\bm u} J_c(\bm u,\bm v^{in})  \\ 
\text{subject to:} \, \bm v^{in} = G_{in}\cdot \bm g_{out}(\bm u,v^{in})  
\end{align}
This formulation avoid the step of designing the local controller at each sub-system's level at the price of using a higher dimensional fixed-point iterating variable that now includes the control profiles. Another advantages of this new framework is that the fixed-point iteration and the solution of the central problem are jointly done and not in two separated phase as in the original settings. 

%Fig. \ref{approach_2}  illustrates the communication between the coordinator and the subsystem.
%\begin{figure}[h]
% \centering
%\includegraphics[width =0.5 \linewidth]{sample-2.pdf}
%\caption{The variables that are exchanged between the coordinator and the subsystems in the new approach.} \label{approach_2}
% \end{figure}
 
This approach has been used for set-point tracking in the case described in Fig. \ref{4_topology}. Figure \label{res_2} shows the behavior of the subsystems using the new approach described above.
\begin{figure}[h]
 \centering
\includegraphics[trim = 1.5cm 4cm 1cm 4.5cm,clip,width = 0.8\linewidth]{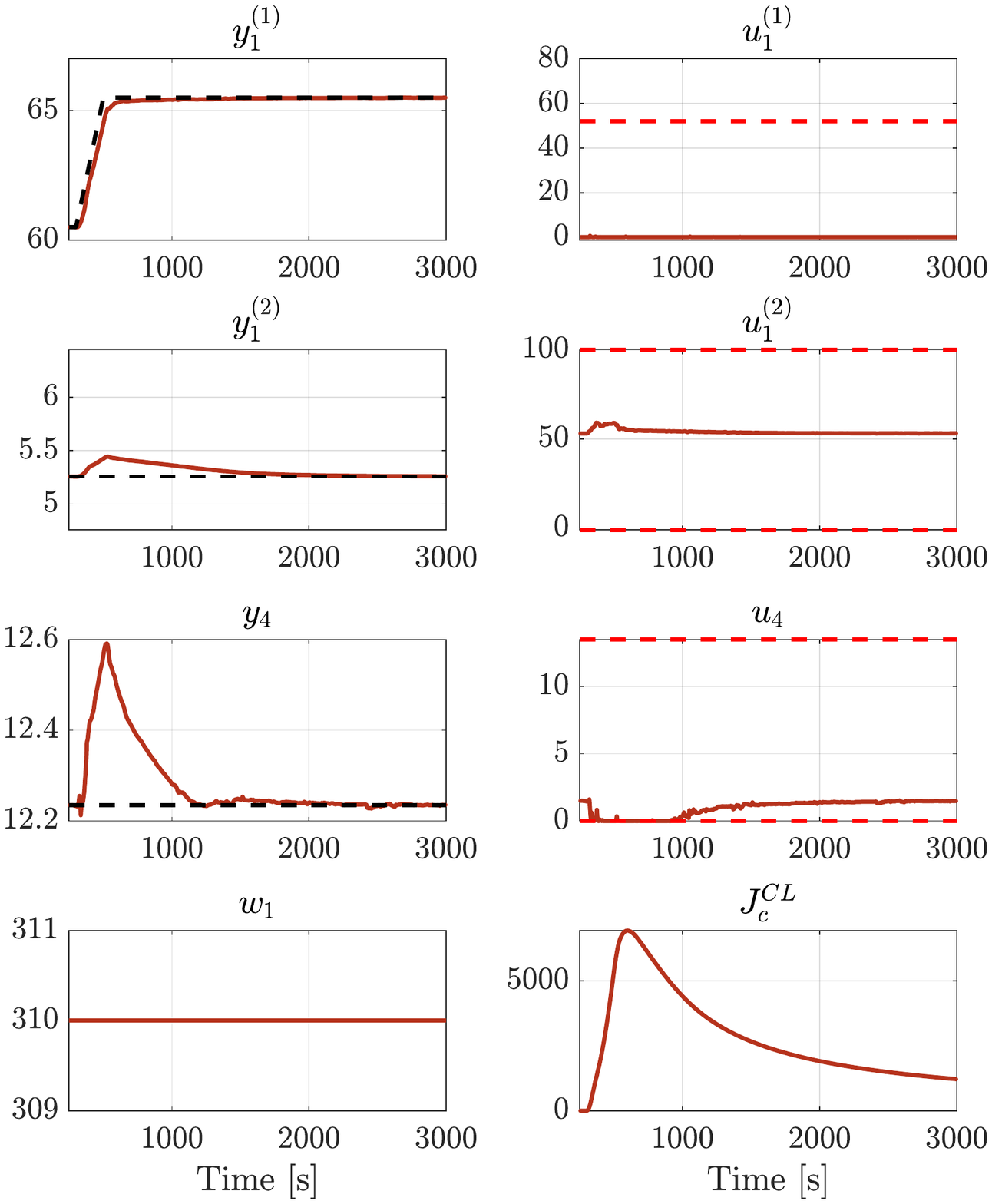}
\caption{Behavior of the network of subsystems depicted in Figure \ref{4_topology} under the hierarchical control described in Section \ref{discus} and using the AA filter. The dash line represents the set-point.} \label{res_2}
 \end{figure}
\section{Conclusion}\label{conclude}
In this paper, the Andersen acceleration filter was applied in the fixed-point updating of a hierarchical control framework.  The method has demonstrated its ability to induce the  convergence of the fixed-point iteration while being totally agnostic to the mathematical model of the subsystems. A new application of the AA method in a modified hierarchical control has been introduced, which gives promising results.
\bibliographystyle{IEEEtran}
\bibliography{IEEEabrv,mybibfile_2}
\end{document}